\documentclass{article}
\usepackage{graphicx}
\usepackage{amsmath}
\usepackage{amsfonts}
\usepackage{amssymb}
\usepackage{epsfig,graphics,graphicx}
\begin{document}

\pagestyle{myheadings} \markright{\it } \vskip.5in
\begin{center}

%
%
\vskip.4in {\Large\bf On the Integrability and Chaos of
an N=2 Maxwell-Chern-Simons-Higgs Mechanical Model} \vskip.3in
%
%
%
L.P.G. de Assis, J.A. Helay\"{e}l-Neto
\footnote{Email: \tt lpgassis@cbpf.br, helayel@cbpf.br, nogue@cbpf.br, ferhaas@unisinos.br}\\
Centro Brasileiro de Pesquisas F\'{\i}sicas - CCP/CBPF - Rio de Janeiro - Brasil {\it and} \\
Grupo de F\'{\i}sica Te\'{o}rica Jos\'{e} Leite Lopes - GFT JLL \\
P.O. Box 91933, 25685-970\\
Petr\'{o}polis  Brasil\\
F. Haas\\
Universidade do Vale do Rio dos Sinos - UNISINOS  \\
Unidade de Exatas e Tecnol\'ogicas \\
Av. Unisinos, 950\\
93022-000 - S\~ao Leopoldo, RS, Brazil\\
A.L.M.A. Nogueira\\
CCP/CBPF, GFT JLL {\it and} Laborat\'{o}rio de Ci\^{e}ncias
F\'{\i}sicas - LCFIS/CCT/UENF

%
%

%
\end{center}
%
\vskip.2in
\begin{abstract}
We apply different integrability analysis procedures to a reduced
(spatially homogeneous) mechanical system derived from an off-shell 
non-minimally coupled N=2 Maxwell-Chern-Simons-Higgs model 
that presents BPS topological vortex excitations,
numerically obtained with an ansatz adopted in a special - critical
coupling - parametric regime. As a counterpart of the regularity
associated to the static soliton-like solution, we
investigate the possibility of chaotic dynamics in the evolution of
the spatially homogeneous reduced system, descendant from the full
N=2 model under consideration. The originally rich content of symmetries and
interactions, N=2 {\it susy} and non-minimal coupling, singles out
the proposed model as an interesting framework for the investigation
of the r\^{o}le played by (super-)symmetries and parametric domains
in the triggering/control of chaotic behavior in gauge systems.

After writing down effective Lagrangian and Hamiltonian functions, and
establishing the corresponding canonical Hamilton equations, we
apply global integrability Noether point symmetries and Painlev\'{e}
property criteria to both the general and the critical coupling
regimes. As a non-integrable character is detected
by the pair of analytical criteria applied, we perform suitable
numerical simulations, as we seek for chaotic patterns in the system
evolution. Finally, we present some Comments on the results and
perspectives for further investigations and forthcoming
communications.
\end{abstract}
 \section{Introduction and Motivation}

The dynamical properties of gauge systems define a focus of great interest,
as one can realize from the remarkable effort that has recently been
driven to the analysis of the stability of gauge field configurations
\cite{chaology}. As a cornerstone of theoretical physics, gauge theories
have been intensively investigated, and the actual result is a comprehensive,
but non-exhaustive, picture of such a successful theoretical framework. One
(promising) aspect of such systems is the room for coherent soliton solutions,
that may play a crucial r\^{o}le in the understanding of physical phenomena
like, for instance, the quark confinement. On the other hand, the search for
chaotic regime windows in gauge theories seems to be as much important as the
former analysis, defining a counterpart approach that may lead to answers to
key-problems, for instance, and again, the confinement phenomenon
\cite{chaology,Chaos-gauge}. In the eighties, a method for investigating chaos in
field theories has been developed and applied to Yang-Mills
systems \cite{Chaos-gauge,savvidy}. The main idea is to reduce the model to
its mechanical limit, by considering {\it spatially homogeneous}
field configurations. The discussion of chaotic evolution in
this restricted regime is conjectured to be sufficient to ensure
chaotic behavior for the full field theory \cite{Chaos-gauge,shur}.

In this context, a relevant focus of attention is defined by the
possibility of establishing a systematic interdependence recipe
for the relation between gauge symmetries and the control of
chaotic dynamics. Among gauge symmetries, one should eventually
care about {\it supersymmetric systems}, considered either as
manifestations of a fundamental symmetry or as enriched models
conceived to be a tool to better describe physical situations.

Planar (2+1) Maxwell-Chern-Simons-Higgs (MCSH) theories, as candidates
for an effective description of high-$T_{c}$ superconductors
phenomena, have recently been chosen as target models to the aim
of order-to-chaos transition studies. Bambah {\it et al.}
\cite{bamba} have considered both the (proven to be) integrable
minimally coupled Chern-Simons-Higgs (CSH) model and its higher
momenta natural extension, namely, the minimally coupled
Maxwell-Chern-Simons-Higgs system. The latter failed when
submitted to an integrability criterium, the Painlev\'{e} test,
leaving room for a chaotic regime that happened to be confirmed by
numerical Lyapunov exponents and phase plots analysis. Recently, 
Escalona {\it et al.} \cite{mexico} have performed a
similar work upon a MCSH system endowed with both minimal {\it and
non-minimal} couplings in the interaction sector. The non-minimal
coupling stands for a Pauli-type term describing a
field-strength/matter-current interaction, admitted in (2+1)-D
regardless of the spin of the matter field \cite{stern}.
Moreover, if quantum extension is a claim, such a
non-minimal coupling should be considered from the start
\cite{kogan}. In the work of Ref.\cite{mexico}, the CSH system is
argued to be still integrable, while the non-minimal MCSH exhibits
``alternating windows of order and chaos", as the non-minimal
coupling constant $g$ is varied, the other parameters defining a
set of constant inputs. The model they adopt is the bosonic
projection of an already established N=1-{\it supersymmetric}
system. As a matter of fact, a non-minimally interacting MCSH system had
formerly deserved an extension endowed with {\it on-shell} N=2-{\it
susy}\cite{navra}. As far as soliton solutions are a subject of
interest, the N=2 extension defines the proper framework, allowing
for the self-dual regime \cite{wit, bogo}. On this token, Antill\'{o}n, Escalona
{\it et al.} have found, in a previous work \cite{antil}, while
working upon such an $N=2$ extendable model, a self-dual {\it static}
non-topological vortex solution, motivating their search for {\it
spatially homogeneous} chaotic dynamics as an interesting
counterpart for their former discovery. Nevertheless, even if one
assumes the validity of the conjecture relating the mechanical
limit to the full theory, a problem arises if the counterpart
character is claimed to be rigorous: the vortex has been found in
an N=2-{\it susy} framework, {\it while the varying $g$ procedure
adopted in \cite{mexico} necessarily moves the system out of the
N=2-{\it susy}-bosonic projection situation}.
As clearly
assumed in \cite{navra}, a {\it critical coupling}, namely, $g =
- e/\kappa$, has to be verified to ensure
on-shell N=2-{\it susy}, where $e$ is the minimal coupling constant and $\kappa$
is the Chern-Simons mass parameter. Moreover, the scalar potential is
forbidden to be anything but the non-topological mass-like
$\phi^{2}$ term. So, varying $g$ while keeping $e$ and $\kappa$
constant, and adopting  $V=\lambda{(\phi^{2} - v^{2})}^{2}$ render
their model a sector of, at most, an N=1 system.

Alternatively, another planar N=2 non-minimal MCSH model has been
recently proposed \cite{numero1}, defining a richer spectrum that
presents both non-topological {\it and topological} self-dual
static vortex solutions \cite{selfdual}, numerically obtained after
the adoption of the critical coupling relation. Such a system exhibits {\it
off-shell-realized N=2-susy}, and is obtained from a N=1-D=4
ansatz, after dimensional reduction and a suitable N=2-covariant
superfield identification. Two important
differences arise, if one settles a comparison encompassing both
non-minimal MCSH models: in the N=2-off-shell case, an
``additional" \footnote{A common, improper, terminology. The
``minimum" content forbids interesting excitations, like
topological vortices.} neutral scalar field takes place; also, in
the N=2-off-shell case, {\it no relation between the coupling
constants and parameters} is required to ensure N=2-{\it susy} (though the
vortex excitations have so far been shown to prevail in the particular
$ g = - e/ \kappa $ regime). In other words, if the model of
Ref.\cite{numero1} is considered, the freely varying $g$ strategy and
a topologically non-trivial scalar potential happen to be compatible with
N=2-{\it susy}.

Motivated by these interesting features, we carry out the analysis of
the reduced - mechanical - version of the bosonic-sector Lagrangian
extracted from the Ref.\cite{selfdual}. In the next section, we
present the theory, the field equations and their spatially
homogeneous counterpart, the one-dimensional effective Lagrangian
and the associated conjugate momenta. As we detect an additional
(besides the Hamiltonian) constant of motion, a convenient
reparametrization is implemented, and the corresponding Hamiltonian
system is displayed, ending up with the canonical Hamilton
equations. In Section 3, we move back to a (general regime)
second-order formulation, upon which we start our analysis of the
integrability issue, adopting two alternative analytical criteria -
the Noether point symmetries approach \cite{sarlet} and the Painlev\'{e} test
procedure \cite{pain1,pain2}. The former strategy leads to a set of ten coupled partial
differential equations which seem to possess a closed form solution
only in the minimal coupling regime, $g = 0 $. In the Painlev\'{e}
test context, it is shown that no set of negative integer dominant
exponents can be found, as the equation for the gauge field is
considered, spoiling the corresponding test algorithm for the
verification of a possible strong Painlev\'{e} property associated
to the system we propose. The r\^{o}le played by the gauge field
equation suggests that a change in the gauge sector dynamics might
render a different situation for the integrability analysis. So
motivated, in Section 4 we adopt the critical coupling relation
regime, arriving at {\it first-order} equations for the gauge
degrees of freedom. We present both the associated effective
Lagrangian and Hamiltonian functions, as well as the corresponding
iterated second-order field equations. These results enable us to
re-address the Painlev\'{e} test, which indicates that the critical
coupling regime presents an even worse feature concerning the
presumed strict negativeness of the dominant exponents. Revisiting
the Noether point symmetries approach also gives no clues on
possible integrable setups. In Section 5, an analysis of chaos is performed
with physically acceptable values of the parameters; in the regimes characterised by
$g\neq0$, for both non-critical and critical values of $g$ , the model becomes more stable
than in the $g=0$ situation. Finally, we present our Concluding Comments and we discuss the
failure of the system in obeying the strong Painlev\'{e}
property, a fact that may lead to future consideration of 
other integrability tests. We also try to explain why $g\neq0$ yields a more stable behavior
in spite of the inherent non-linearity brought about by a non-vanishing $g$. 

\section{Describing the Model}
We start \cite{selfdual} with:
\begin{eqnarray}
{\cal L}_{\mbox{\tiny boson}} & = & -\frac{1}{4}F_{\mu \nu }^{2}
+\frac{1}{2}\partial_{\mu} M\partial^{\mu} M +
\frac{1}{2}(\nabla_{\mu} \phi){( \nabla^{\mu} \phi)}^{*} +
\nonumber \\
&& - \frac{g}{2} (\partial_{\mu}M)(\partial^{\mu}{|\phi|}^{2}) +
\frac{\kappa}{2} A_{\mu} {\tilde{F}}^{\mu} - U , \label{T-U}
\end{eqnarray}

\noindent where
\begin{eqnarray}
U & = & \frac{e^2}{8G}{\left( |\phi |^{2} - v^{2} +\frac{2m}{e}M +
2g|\phi |^{2}M\right)}^{2} + \frac{e^{2}}{2}M^{2}|\phi |^{2}\; ,
\nonumber
\end{eqnarray}

\noindent and $\nabla_{\mu}\phi \;\equiv\; (\partial_{\mu}
-ieA_{\mu} -ig {\tilde{F}}_{\mu})\phi$. $G$ is defined as $G\equiv
1 - g^{2}{|\phi|}^{2}$. The field equations for the full theory
read:
\begin{eqnarray}
\partial_{\mu} F^{\mu \rho }+ m {\tilde{F}}^{\rho} & = & -{\cal J}^{\;\rho }-\frac{g}{e}
{\varepsilon}^{\mu \nu \rho }\partial_{\mu} {\cal J}_{\nu}\; ,
\nonumber
\end{eqnarray}
\noindent where $ {\cal J}_{\mu} \; = \;
\frac{ie}{2}({\phi}^{*}\nabla_{\mu}\phi -
\phi{(\nabla_{\mu}\phi)}^{*})$, and also

\begin{eqnarray}
\partial_{\alpha}\partial^{\alpha}M
 -  \frac{g}{2}\partial_{\alpha}\partial^{\alpha}{|\phi|}^{2}
+ \frac {e^{2}({|\phi|}^{2} - v^{2} + (2\kappa/e)M +
2gM{|\phi|}^{2})(2\kappa/e + 2g{|\phi|}^{2})}{4G}
+  e^{2}{|\phi|}^{2}M  =  0 &&\; , \nonumber
\end{eqnarray}
\begin{eqnarray}
{\mbox {\small and}}\;\;\;\;\frac{1}{2}{(\nabla_{\alpha}\nabla^{\alpha}\phi)}^{*}  -
\frac{g}{2} \phi^{*} (\partial_{\alpha}\partial^{\alpha}M)
+\frac {e^{2}g^{2}\phi^{*}{\left[ {|\phi|}^{2} -v^{2} +
(2\kappa/e)M + 2g {|\phi|}^{2}M \right]}^{2}}{8G^{2}} + &  &
\nonumber \\ + \frac {e^{2}\phi^{*}\left[ {|\phi|}^{2} -v^{2} +
(2\kappa/e)M + 2g
{|\phi|}^{2}M \right](1 + 2gM)}{4G}
+ \frac{e^{2}\phi^{*}M^{2}}{2}  =  0 &&\nonumber\; .
\end{eqnarray}

Adopting the gauge choice $A_{0} = 0$ and imposing the spatial
homogeneity, namely, $\partial_{i}(\forall \; {\mbox field})\; =
0$, the phase of the scalar field becomes a variable with vanishing time-derivative,
and one can eliminate it without loss of generality. So,
one ends up with a real scalar field and the following set of reduced equations of motion:

\begin{eqnarray}
\frac{d}{dt}\left[G{\dot{A}}_{1}\right] & = & - 2eg\phi
\frac{d}{dt}\left[\phi A_{2}\right] -e^{2}\phi^{2}A_{1} - \kappa
{\dot{A}}_{2} \label{a1} \; ,
\end{eqnarray}
\begin{eqnarray}
\frac{d}{dt}\left[G{\dot{A}}_{2}\right] & = & + 2eg\phi
\frac{d}{dt}\left[\phi A_{1}\right] - e^{2}\phi^{2}A_{2} + \kappa
{\dot{A}}_{1} \label{a2} \; ,
\end{eqnarray}
\begin{eqnarray}
\ddot{M}  =  \frac{g}{2}\dot{(\phi^{2})} -  e^{2} \phi^{2} M
- \frac {e^{2}({\phi}^{2} - v^{2} + (2\kappa/e)M +
2gM{\phi}^{2})(2\kappa/e + 2g{\phi}^{2})}{4G} && \nonumber \;
,
\end{eqnarray}
\begin{eqnarray}
\ddot{\phi} = -{(eA_{1} +g {\dot{A}}_{2})}^{2}\phi -{(eA_{2} -g
{\dot{A}}_{1})}^{2}\phi + g\phi\ddot{M} -e^{2} \phi M^{2} && \nonumber \\
- \frac {e^{2}g^{2}\phi{({\phi}^{2} - v^{2} + (2\kappa/e)M +
2gM{\phi}^{2})}^{2}}{4G^{2}} && + \nonumber \\
- \frac {e^{2}\phi({\phi}^{2} - v^{2} + (2\kappa/e)M +
2gM{\phi}^{2})(1 + 2gM)}{2G} && \nonumber  \; .
\end{eqnarray}

The effective Lagrangian that generates these equations of motion is:

\begin{eqnarray}
L  =  \frac{G}{2} \left[ {({\dot{A}}_{1})}^{2} +
{({\dot{A}}_{2})}^{2}\right] - \frac{Q}{2}\left(
A_{1}{\dot{A}}_{2} - A_{2}{\dot{A}}_{1}\right)
 -\frac{e^{2}\phi^{2}}{2}\left( {A_{1}}^{2} + {A_{2}}^{2} \right)+
&&\nonumber\\
+ \frac{{(\dot{M})}^{2}}{2} - g\phi\dot{\phi}\dot{M}
+ \frac{{(\dot{\phi})}^{2}}{2}
- \frac {e^{2}{({\phi}^{2} - v^{2} + (2\kappa/e)M +
2gM{\phi}^{2})}^{2}}{8G}
- \frac{e^{2}\phi^{2}M^{2}}{2}
&&\nonumber \; ,
\end{eqnarray}

\noindent where $G\equiv 1 - g^{2}\phi^{2}$ and $Q\equiv \kappa +
2eg\phi^{2}$.

The canonically conjugate momenta, defined, as usually, by $p =
\frac{\partial L}{\partial \dot{q}}$, have the expressions:
\begin{eqnarray}
\pi_{1} \equiv \frac{\partial L}{\partial {\dot{A}}_{1}} & = & G
{\dot{A}}_{1} + \frac{Q}{2}A_{2} \; ,\nonumber \\
\pi_{2} \equiv \frac{\partial L}{\partial {\dot{A}}_{2}} & = & G
{\dot{A}}_{2} - \frac{Q}{2}A_{1}  \; , \nonumber \\
p_{\phi} \equiv \frac{\partial L}{\partial \dot{\phi}} & = &
\dot{\phi} -g\phi\dot{M} \; , \nonumber \\
P_{M} \equiv \frac{\partial L}{\partial \dot{M}} & = & \dot{M} -
g\phi\dot{\phi}\; . \nonumber
\end{eqnarray}

Before we proceed to the canonical Hamilton equations, let us
notice that the quantity
\begin{eqnarray}
I &\equiv & A_{2}\pi_{1} - A_{1}\pi_{2} \label{const}
\end{eqnarray}
\noindent is a constant of motion. This can easily be checked
by multiplying and combining the ``gauge fields" equations
according to $A_{2}(\ref{a1}) - A_{1}(\ref{a2})$. Motivated by
this fact we reparametrize the gauge sector adopting polar
coordinates, instead of Cartesian ones. We have: $A_{1} = A
cos\zeta$, $A_{2} = A sen\zeta $, and the ``new" set of variables
is $(A,\zeta,\phi,M)$. The Lagrangian now reads:
\begin{eqnarray}
L = \frac{G}{2}\left[ {(\dot{A})}^{2} +
A^{2}{(\dot{\zeta})}^{2}\right]
-\frac{Q}{2}\left(A^{2}\dot{\zeta}\right)
+ \frac{{(\dot{M})}^{2}}{2} - g\phi\dot{\phi}\dot{M}
+ \frac{{(\dot{\phi})}^{2}}{2} &&\nonumber\\
-\frac{e^{2}\phi^{2}A^{2}}{2} - \frac {e^{2}{({\phi}^{2} - v^{2} +
(2\kappa/e)M + 2gM{\phi}^{2})}^{2}}{8G} -
\frac{e^{2}\phi^{2}M^{2}}{2} && \label{lagran} \; ,
\end{eqnarray}
\noindent yielding the same expressions for $p_{\phi}$ and
$P_{M}$, and defining $p_{A}\equiv\frac{\partial L}{\partial
\dot{A}} = G\dot{A}$ , $p_{\zeta}\equiv\frac{\partial L}{\partial
\dot{\zeta}} = GA^{2}\dot{\zeta} - \frac{Q}{2} A^{2}$. One can
easily check that $p_{\zeta} = - I $, resulting ${\dot{p}}_{\zeta}
= 0$. The Hamiltonian reads:

\begin{eqnarray}
H_{CAN.} = \frac{1}{2G}\left[ {p_{A}}^{2} +
\frac{{p_{\zeta}}^{2}}{A^{2}} + Q p_{\zeta} + {p_{\phi}}^{2} +
{P_{M}}^{2} + 2g\phi p_{\phi} P_{M} \right] + && \nonumber \\
+ \frac{1}{2G} \left[ {(Q/2)}^{2} + e^{2} G\phi^{2}\right] A^{2} +
\frac{e^{2}\phi^{2}M^{2}}{2}
+ \frac{e^{2}}{8G}{\left( \phi^{2} - v^{2} + (2\kappa/e)M +
2g\phi^{2}M \right)}^{2} &&\; .\nonumber
\end{eqnarray}

The canonical Hamilton equations result:
\begin{eqnarray}
\dot{A} \; = \; \frac{p_{A}}{G}\; \; & ; & \;\;
{\dot{p}}_{A} \; = \;\frac{1}{G} \left[
\frac{{p_{\zeta}}^{2}}{A^{3}} - \left( {(Q/2)}^{2} + e^{2}
G\phi^{2}\right) A \right]\; ; \nonumber
\end{eqnarray}
\begin{eqnarray}
\dot{\zeta} & = & \frac{1}{G} \left[ \frac{p_{\zeta}}{A^{2}} +
\frac{Q}{2} \right]\; ; \; \; {\dot{p}}_{\zeta} = 0 \; ;
\label{zeta}
\end{eqnarray}
\begin{eqnarray}
\dot{\phi} & = & \frac{1}{G}\left[ p_{\phi} + g\phi P_{M} \right]
\; ; \nonumber
\end{eqnarray}
\begin{eqnarray}
{\dot{p}}_{\phi} = -\frac{1}{G^{2}}\left\{ g^{2}\phi \left[
{p_{A}}^{2} + \frac{{p_{\zeta}}^{2}}{A^{2}} + {p_{\phi}}^{2} +
{P_{M}}^{2} \right]
+g\phi(\kappa g + 2e)p_{\zeta} + g (1+g^{2}\phi^{2})p_{\phi}P_{M}
\right. && \nonumber \\ + {(\kappa g + 2e)}^{2} \frac{\phi A^{2}}{4} +
e^{2}G^{2}\phi M^{2}
+ \frac{e^{2}g^{2}\phi}{4}{\left( \phi^{2} -v^{2} + (2\kappa/e)M +
2gM\phi^{2} \right)}^{2} && \nonumber \\
+ \left. \frac{e^{2}G\phi}{2}\left( \phi^{2} -v^{2} + (2\kappa/e)M
+ 2gM\phi^{2} \right)(1 + 2gM)\right\} && \; ;\nonumber
\end{eqnarray}
\begin{eqnarray}
\dot{M}& = & \frac{1}{G}\left[ P_{M} + g\phi p_{\phi} \right]\; ;
\nonumber
\end{eqnarray}
\begin{eqnarray}
{\dot{P}}_{M} = -\frac{1}{G}\left[ e^{2}G\phi^{2} M
+ \frac{e^{2}}{4}\left( \phi^{2} - v^{2} + (2\kappa/e)M +
2gM\phi^{2} \right)\left( \frac{2\kappa}{e} + 2g\phi^{2} \right)
\right] && \; . \nonumber
\end{eqnarray}

\section{Integrability Analysis: General Case}

We present two Lagrangian analytical criteria to address the
issue of integrability: Noether point symmetries, better suited for
establishing the constants of motion, and the Painlev\'{e} test,
meant to check for an overall property (the dependent variables
being meromorphic for movable singularities on the complex time
plane) that indicates integrability.

\subsection{Noether point symmetries}

An important issue regarding a Lagrangian system concerns its
Noether point symmetries, linking symmetries of the action
functional to conserved quantities. Here we address the question of
the existence of Noether point symmetries of our system following
the method shown in reference \cite{sarlet}.

We seek for infinitesimal point transformations of the form
\begin{eqnarray}
\bar{A} &=& A + \varepsilon\eta_A \,,\\
\bar{\zeta} &=& \zeta + \varepsilon\eta_\zeta \,,\\
\bar{\phi} &=& \phi + \varepsilon\eta_\phi \,,\\
\bar{M} &=& M + \varepsilon\eta_M \,,\\
\bar{t} &=& t + \varepsilon\tau \,,
\end{eqnarray}
for $\eta_{A}, \eta_{\zeta}, \eta_{M}, \eta_{\phi}$ and $\tau$
functions of the fields and the time, and $\varepsilon$ an
infinitesimal parameter. These infinitesimal transformations leave
the action functional invariant up to the addition of an irrelevant
numerical constant if and only if the following Noether symmetry
condition \cite{sarlet} is satisfied,
\begin{eqnarray}
\tau\frac{\partial L}{\partial t} &+& \eta_{A}\frac{\partial L}{\partial A} + \eta_{\zeta}\frac{\partial L}{\partial\zeta} + \eta_{\phi}\frac{\partial L}{\partial\phi} + \eta_{M}\frac{\partial L}{\partial M} + (\dot{\eta}_A - \dot{\tau}\dot{A})\frac{\partial L}{\partial\dot{A}} + \nonumber \\
&+& (\dot{\eta}_\zeta - \dot{\tau}\dot{\zeta})\frac{\partial
L}{\partial\dot{\zeta}} + (\dot{\eta}_\phi -
\dot{\tau}\dot{\phi})\frac{\partial L}{\partial\dot{\phi}} +
(\dot{\eta}_M - \dot{\tau}\dot{M})\frac{\partial L}{\partial\dot{M}} + \nonumber \\
&+& \dot\tau\,L = \dot{F} \,.
\end{eqnarray}
for $F$ a function of the fields and the time. If such a function
can be found, there is a Noether point symmetry and an associated
Noether invariant $I$ given by
\begin{equation}
\label{inva} I = \eta_{A}\frac{\partial L}{\partial\dot{A}} +
\eta_{\zeta}\frac{\partial L}{\partial\dot{\zeta}} +
\eta_{\phi}\frac{\partial L}{\partial\dot{\phi}} +
\eta_{M}\frac{\partial L}{\partial\dot{M}} -
\tau(\dot{A}\frac{\partial L}{\partial\dot{A}} +
\dot{\zeta}\frac{\partial L}{\partial\dot{\zeta}} +
\dot{\phi}\frac{\partial L}{\partial\dot{\phi}} +
\dot{M}\frac{\partial L}{\partial\dot{M}} - L) - F \,.
\end{equation}

In the Noether symmetry condition, the time derivatives are to be
understood as total derivatives, e.g.,
\begin{equation}
\dot\tau = \frac{\partial\tau}{\partial A}\dot{A} +
\frac{\partial\tau}{\partial\zeta}\dot{\zeta} +
\frac{\partial\tau}{\partial\phi}\dot{\phi} +
\frac{\partial\tau}{\partial M}\dot{M} +
\frac{\partial\tau}{\partial t} \,.
\end{equation}

Inserting in the symmetry condition the Lagrangian given by equation
(\ref{lagran}), we obtain that a cubic polynomial in the velocities
must vanish. The coefficients of equal powers of velocities
vanishing, we obtain a coupled set of linear partial differential
equations determining both the symmetries and the Noether
invariants. The cubic terms yields simply
\begin{equation}
\tau = \tau(t) \,,
\end{equation}
that is, the transformed independent variable is a function of time
only. The equations associated to quadratic terms, however, are a
set of ten coupled partial differential equations which, apparently,
do possess a closed form solution only in the minimal coupling case
$g = 0$. Restricting the treatment to this almost trivial case, and
proceeding to the first and zeroth order terms, we just found
time-translation and $\zeta$ translation symmetries. These
symmetries are associated, according to (\ref{inva}), to the energy
and $p_\zeta$ conservation laws. These are almost obvious results,
showing that the nonlinearity and coupling in the potential gives no
much space for the existence of conservation laws of the system,
even in the $g = 0$ case. This is a signature of  non-integrability.
However, other methods for investigating conservation laws of the
system like Lie point symmetries \cite{olver} were not used.

\subsection{Painlev\'{e} test}
We now go back to a second-order configuration space
formalism in order to settle the framework for the application of
Painlev\'{e} test \cite{pain1,pain2}. The equations read

\begin{eqnarray}
G^{2}\ddot{A} & = & 2g^{2}G\phi\dot{\phi}\dot{A} +
\frac{C^{2}}{A^{3}}
-{(\frac{\kappa}{2})}^{2} A - e (\kappa g + e)  \phi^{2} A
\label{painleve1} \; ,
\end{eqnarray}
\noindent where $C$ stands for the constant $p_{\zeta}$,
%
\begin{eqnarray}
G^{3}\ddot\phi &=& g^{2}G^{2}\phi\dot\phi^2 - g^{2}G^{2}\phi\dot{A}^2 - \frac{g^{2}C^{2}\phi}{A^2} - gC(\kappa\,g + 2e)\phi - \frac{1}{4}(\kappa\,g + 2e)^{2}\phi\,A^2 + \nonumber \\ &+& \frac{e\phi}{4}(\phi^2 - v^2)\left[eg^{2}v^2 - 2(\kappa\,g + e) + g^{2}(2\kappa\,g - e)\phi^2 + 2eg^{4}\phi^4\right] \nonumber \\
&-& (\kappa\,g + e)\left[\kappa - ev^2\,g + (3eg - \kappa\,g^2)\phi^2 - 2eg^3\phi^4 + (\kappa\,g + e)M\right]\phi\,M   \label{painleve2}
\end{eqnarray}
and
\begin{eqnarray}
G^{3}\ddot{M} &=& gG^{2}{\dot\phi}^2 - g^{3}G^{2}\phi^{2}{\dot A}^2 + \frac{g^{3}C^{2}\phi^2}{A^2} - \frac{g}{4}(\kappa\,g + 2e)^{2}A^{2}\phi^2 - g\,(\kappa\,g + e)^{2}M^{2}\phi^{2} + \nonumber \\
&+& \frac{e}{4}(\phi^2 - v^2)[- 2\kappa + g\,(- 4e + eg^{2}v^2 + 2\kappa\,g)\phi^2 + 3eg^{3}\phi^4] \nonumber \\
&+& [- \kappa^2 + (\kappa^{2}g^2 + e\kappa\,g^{3}v^{2} - e^2 - 3e\kappa\,g + e^{2}g^{2}v^{2})\phi^2 + eg^{2}(2\kappa\,g - e)\phi^4 + e^{2}g^{4}\phi^{6}]M \,. \label{painleve3}
\end{eqnarray}

Assuming time to be a complex variable, the first step of the
Painlev\'{e} test is concerned with the leading singularity
behavior. One supposes the leading terms to be of the general
form $A \sim a{(t - t_{0})}^{\alpha}, \phi \sim b{(t -
t_{0})}^{\beta}, M \sim c{(t - t_{0})}^{\gamma}$, where $\alpha ,
\beta , \gamma < 0$. Such an assumption turns the last three equations into
the following asymptotic ($ t \rightarrow t_{0} $) relations:

\begin{eqnarray}
g^{4}a b^{4}\alpha \left[ \alpha -1 + 2\beta \right]{\tau}^{\alpha + 4\beta -2}
& \sim & 0 \;\; ;\nonumber
\end{eqnarray}
\begin{eqnarray}
g^{6}b^{7}\beta \left[ 2\beta -1 \right]{\tau}^{7\beta -2} & \sim &
g^{6}b^{5}a^{2}{\alpha}^{2}{\tau}^{2\alpha + 5\beta -2} \; + \nonumber \\
-2eg^{3}(\kappa g + e) b^{5} c {\tau}^{5\beta + \gamma} & + &
{(\kappa g + e)}^{2} b c^{2} {\tau}^{\beta + 2\gamma} \;\; ; \nonumber
\end{eqnarray}
\begin{eqnarray}
c \gamma (\gamma -1) {\tau}^{4\beta + \gamma -2} & \sim &
gb^{2}\beta (2\beta - 1){\tau}^{6\beta -2} \;\; , \nonumber
\end{eqnarray}
\noindent where $\tau = t - t_{0}$.

Starting from the last equation, one gets $\gamma = 2\beta $ and
$ c = {gb^{2}}/{2}$. Inserting $\gamma = 2\beta $ in the second equation
reduces the balancing to the first two terms, so leading to $\alpha = \beta $
and $ a^{2} = {(2\beta -1)b^{2}}/{\beta}$. But the first equation shows the
impossibility of having $\alpha , \beta < 0 $, as $\alpha + 2\beta -1 = 0$ is
required, spoiling the Painlev\'{e} test procedure.

Another possibility would be to set $\alpha = 0 $ in the first
equation, leaving it behind as an identity. One could then drop the
second term (first on right-hand side) of the second equation, and
the balancing of the remaining three terms would lead to an
interesting set of negative values for $\gamma$ and $\beta$: $\gamma
= -4, \, \beta = -1$, provided that the following relation holds:

\begin{eqnarray}
{(\kappa g + e )}^{2} c^{2} - 2 (\kappa g + e ) e g^{3}b^{4}c +
3g^{6}b^{6} & = & 0  \nonumber\; .
\end{eqnarray}

Still, one has to deal with a zero ``dominant" exponent, which
spoils the Painlev\'{e} test.

As one faces a problem with the gauge sector dynamics, the adoption
of the critical coupling relation (projecting the system onto the
regime that hosts the already established vortex excitations) may
serve as a valuable tool of investigation. In fact, imposing $g =
-{e}/{\kappa}$ leads to first-order equations for the gauge field
\cite{selfdual}.

\section{Critical coupling regime}

If $g = -{e}/{\kappa}$, one gets:
\begin{eqnarray}
\kappa {\tilde{F}}_{\nu} & = & - {\cal J}_{\nu} \;\; , \nonumber
\end{eqnarray}
\noindent and the reduction to spatially homogeneous configurations yields
\begin{eqnarray}
\kappa  G {\cal E}_{ij}\, {\dot{A}}_{j} & = & - e^{2} \, A_{i}\, {\phi}^{2}
\;\; , \label{ordemuno}
\end{eqnarray}
\noindent where ${\cal E}_{12} = + 1 = - {\cal E}_{21}$.
From this set of equations one can arrive at
\begin{eqnarray}
G \frac{d}{dt}(A_{1}^{2} + A_{2}^{2}) & = & 0 \;\; , \nonumber
\end{eqnarray}
\noindent and, as far as $ G > 0$ (a condition inherited from the
original N=2-{\it susy} framework), this implies that $A_{1}^{2} +
A_{2}^{2}$ is a constant of motion (thus reproducing the ``pure"
minimally coupled Chern-Simons-Higgs situation). Adopting polar
coordinates, $A_{1} = C cos\,\zeta$, $A_{2} = C sen \, \zeta$
($A_{1}^{2} + A_{2}^{2} = C^{2}$), and manipulating the set
(\ref{ordemuno}), one finds $\dot{\zeta}\, = \, -
{e}^{2}{\phi}^{2}/{\kappa} \, G $. Following the same route chosen
in the general (non-critical) case, we seek for the effective
Lagrangian and Hamiltonian, settle the canonical equations of motion
and, as we aim at the Painlev\'{e} test for integrability, iterate
them to get second-order coupled differential equations. One can
easily verify that the following Lagrangian and Hamiltonian
functions

\begin{eqnarray}
L & = & p_{\zeta}\left( \dot{\zeta} +
\frac{{e}^{2}{\phi}^{2}}{\kappa G}\right) +
\frac{{\dot{\phi}}^{2}}{2} - \frac{{e}^{2}C^{2}{\phi}^{2}}{2} -
\frac{{e}^{4}C^{2}{\phi}^{4}}{2{\kappa}^{2}G} + \nonumber \\
&& \frac{{\dot{M}}^{2}}{2} - g\phi\dot{\phi}\dot{M} -
\frac{{e}^{2}{\left( \phi^{2} - v^{2} + (2\kappa/e)M + 2g\phi^{2}M
\right)}^{2}}{8G} - \frac{{e}^{2}M^{2}{\phi}^{2}}{2}\;\; , \nonumber
\end{eqnarray}

\begin{eqnarray}
H_{CAN.} = \frac{1}{2G}\left[
\frac{{p_{\zeta}}^{2}}{A^{2}} + Q p_{\zeta} + {p_{\phi}}^{2} +
{P_{M}}^{2} + 2g\phi p_{\phi} P_{M} \right] + && \nonumber \\
+ \frac{1}{2G} \left[ {(Q/2)}^{2} + e^{2} G\phi^{2}\right] A^{2} +
\frac{e^{2}\phi^{2}M^{2}}{2}
+ \frac{e^{2}}{8G}{\left( \phi^{2} - v^{2} + (2\kappa/e)M +
2g\phi^{2}M \right)}^{2} && \; , \nonumber
\end{eqnarray}
\noindent where $g = -{e}/{\kappa}$, $G = 1 - (e^{2}/{\kappa}^{2}){\phi}^{2}$,
and $p_{\zeta} = - {\kappa}{C}^{2}/2 $ \footnote{In fact, such a relation
between the conserved quantities $p_{\zeta}$ and
${A}^{2} \, (\equiv {C}^{2})$ must be
imposed to ensure first-order
$\dot{\zeta} \, = \, - e^{2}{\phi}^{2} /{\kappa} G $
equation of motion.}, leads to the proper set of
field equations. The iterated second-order system turns out to be

\begin{eqnarray}
-G^{3}\ddot{\phi} = -g^{2}{G}^{2}\phi{\dot{\phi}}^{2}
+\frac{g^{2}C^{2}\phi}{A^{2}}
+ {e}^{2}\frac{\phi A^{2}}{4} && \nonumber \\ + \left[
eg (C + egv^{4}/4)\right]
\phi \;
+\frac{e}{2}\left[ -2eg^{2}v^{2}\right]\phi^{3} + e^{2}g^{2}\left[1 +
\frac{g^{2}v^{2}}{4}\right] \phi^{5} -\frac{e^{2}g^{4}}{2}\phi^{7}\; ;
&& \nonumber
\end{eqnarray}
\begin{eqnarray}
G^{2}\ddot{M} = (g - 2g^{3}\phi^{2} + g^{5}\phi^{4})
{\dot{\phi}}^{2}
+ (g\phi -2g^{3}\phi^{3} + g^{5}\phi^{5})\ddot{\phi} +
2e^{2}\phi^{2}M && \nonumber \\
- e^{2}g^{2}\phi^{4}M -\kappa^{2}M
- \frac{\kappa e}{2}\phi^{2}
- e^{2}g^{2}(1 + \frac{g^{2}v^{2}}{2})\phi^{4}
+ \frac{e^{2}g^{3}\phi^{6}}{2} + \frac{\kappa e v^{2}}{2} \; . &&
\nonumber
\end{eqnarray}

Again, for the Painlev\'{e} test, the asymptotic relations are
found: $\phi: \; \beta = 0$ or $\beta = 1/2 $. If one takes the
$\beta = 0$ case, one is left with two problematic outputs, as the
equation for M is considered: either $\gamma = \beta = 0$, or $\beta
= 0$, $b^{2} = 1/{g}^{2}$, $\gamma $ undetermined. So the signature
of lack of strong Painlev\'{e} property remains.


\section{\bigskip Analysis of chaos}

Since the results of the analytical approaches suggest that the system may not
be integrable, we now turn to a numerical study to verify if such a
non-integrability feature is presented in a chaotic form.

\subsection{SALI method}

The most well--known method used to detect whether a system is chaotic or not is
the maximal Lyapunov Characteristic Exponent (LCE), $\sigma_{1}$. If
$\sigma_{1}>0$ the flow is chaotic. The $\sigma_{1}$ is computed \cite{lya1,lya2} from
\begin{equation}
L_{t}=\frac{1}{t}\,\ln\frac{|\vec{w}(t)|}{|\vec{w}(0)|}\,,\,\mbox
{performing the limit}\,\,\label{eq:lyap1}%
\end{equation}%

\begin{equation}
\sigma_{1}=\lim_{t\rightarrow\infty}L_{t}\,,\label{eq:lyap2}%
\end{equation}
where $\vec{w}(0)$, $\vec{w}(t)$ are deviation vectors and the time evolution of
$\vec{w}$ is given by solving the \textit{equations of motion} and associated \textit{variational equations}.

Since these vectors tend to acquire an exponential growth in short time
intervals, many calculations of $L_{T_{1}}$, as  $\vec{w}(t)$ evolves for a short time $t_{1}$, are carried out after each $\vec{w}(t)$ is normalizated. With this procedure,
the mean value of $L_{T_{1}}$ is computated as
\[
\sigma_{1}=\frac{1}{N}\sum_{i=1}^{N}L_{T_{i}}%
\]

\bigskip For Hamiltonian sytems, this computation becomes very lengthy with
poor convergence, and this long procedure may point to a false chaos diagnostic.

We have chosen to adopt the method developed by Skokos, Antonopoulos, Boutis
and Vrahatis, the so-called Smaller Alignment Index, SALI, for brevity\cite{SABV1,SABV2}. The
reason for this choice is that the SALI method is computationally faster and
less unstable than the Lyapunov exponent analysis,  improving the adequacy of
the former for the system we investigate. The SALI is a indicator of chaos that
tends to zero for chaotic orbits, while it exhibits small fluctuations around
non-zero values for ordered ones.  So, the SALI is defined as:
\begin{equation}
\mbox{SALI}(t)= \min\left\{  \left\|  \frac{\vec{w}_{1}(t)}{\|\vec{w}%
_{1}(t)\|}+ \frac{\vec{w}_{2}(t)}{\|\vec{w}_{2}(t)\|} \right\|  , \left\|
\frac{\vec{w}_{1}(t)}{\|\vec{w}_{1}(t)\|} -\frac{\vec{w}_{2}(t)}{\|\vec{w}%
_{2}(t)\|} \right\|  \right\} ,\label{eq:SALI}%
\end{equation}
where $\vec{w}_{1}(t)$ and $\vec{w}_{2}(t)$ are the evolutions of two deviations vectors with different initial conditions, $\|\cdot\|$ is the Euclidean norm and $t$ is the time . 

The authors of SALI method showed that SALI can be approximated by means of the difference of the two largest Lyapunov
characteristic exponents $\sigma_{1}$ and $\sigma_{2}$.

The main advantage of the SALI in chaotic regions is that it uses two
deviation vectors and exploits at every step the convergence from all previous
steps. The SALI value tends to zero for chaotic flows
at a rate which is a function of the difference of the two largest Lyapunov
characteristic exponents $\sigma_{1}$, $\sigma_{2}$ as $\mbox{SALI}\propto
e^{-(\sigma_{1}-\sigma_{2})t}$. As usually done in numerical computations, we need to define a threshold so that a
computed number be considered zero. In most of the cases, the selected value is $< 10^{-5}$. Like in the case of the Lyapunov exponent, it also happens that, in the SALI method, this is the criterion we shall use to distinguish between order and chaos.

\bigskip

\subsection{Equations of motion and requirements}

The integration of the system was performed by means of a Gear algorithm, in a
variable step mode, starting from the minimal step size h = 0.0001, and
eventually getting reduced in order to preserve the value of the Hamiltonian
and $p_{\zeta}$, known to be constants of motion. Another constraint
maintained along the integration was that G ($G\equiv1-g^{2}\phi^{2}$) should
be greater than zero. The following first order equations of
motion were used in the numerical integration:%

\[
\dot{A}\;=\;\frac{p_{A}}{G}\;\;;\;\;{\dot{p}}_{A}\;=\;\frac{1}{G}\left[
\frac{{p_{\zeta}}^{2}}{A^{3}}-\left(  {(Q/2)}^{2}+e^{2}G\phi^{2}\right)
A\right]  \;;
\]%
\[
\dot{\zeta}=\frac{1}{G}\left[  \frac{p_{\zeta}}{A^{2}}+\frac{Q}{2}\right]
\;;\;\;{\dot{p}}_{\zeta}=0\;;
\]%
\[
\dot{\phi}=\frac{1}{G}\left[  p_{\phi}+g\phi P_{M}\right]  \;;
\]%
\begin{align}
{\dot{p}}_{\phi}=-\frac{1}{G^{2}}\left\{  g^{2}\phi\left[  {p_{A}}^{2}%
+\frac{{p_{\zeta}}^{2}}{A^{2}}+{p_{\phi}}^{2}+{P_{M}}^{2}\right]
+g\phi(\kappa g+2e)p_{\zeta}+g(1+g^{2}\phi^{2})p_{\phi}P_{M}\right.   &
\nonumber\\
+{(\kappa g+2e)}^{2}\frac{\phi A^{2}}{4}+e^{2}G^{2}\phi M^{2}+\frac{e^{2}%
g^{2}\phi}{4}{\left(  \phi^{2}-v^{2}+(2\kappa/e)M+2gM\phi^{2}\right)  }^{2}  &
\nonumber\\
+\left.  \frac{e^{2}G\phi}{2}\left(  \phi^{2}-v^{2}+(2\kappa/e)M+2gM\phi
^{2}\right)  (1+2gM)\right\}   &  \;;\nonumber
\end{align}%
\[
\dot{M}=\frac{1}{G}\left[  P_{M}+g\phi p_{\phi}\right]  \;;
\]%
\[
{\dot{P}}_{M}=-\frac{1}{G}\left[  e^{2}G\phi^{2}M+\frac{e^{2}}{4}\left(
\phi^{2}-v^{2}+(2\kappa/e)M+2gM\phi^{2}\right)  \left(  \frac{2\kappa}%
{e}+2g\phi^{2}\right)  \right]  \;.
\]

\bigskip For the sake of simplicity, we have adopted the following notation:

$\zeta=q_{1},\ A=q_{2},\ \phi=q_{3},\ M=q_{4},\ p_{\zeta}=p_{1},p_{A}%
=p_{2},\ p_{\phi}=p_{3},\ P_{M}=p_{4}.$

\bigskip For each set of parametric inputs the numerical integration was
performed and SALI method used after transient damped. In the following, we display representative samples of our findings.

\bigskip

\subsection{\bigskip Case with g=0}

\bigskip

\bigskip  Since our model comes from a supersymmetric version of
Maxwell-Chern-Simons-Higgs with non-minimal coupling, it is not clear whether we
shall recover dynamical properties similar to the ones observed in other studies \cite{mexico,bamba}
when $g=0$ is adopted, and the model is so reduced to the minimal coupling case. The two main
properties found in these studies were the existence of chaos in the
presence of Maxwell term and the asymptotic evolution - order versus chaos - sensibility to initial
conditions. To verify these properties in our model, we have chosen the
initial conditions defining a fixed point of the system, and then we varied $q_{3}$
from $0$ to $2$ with parameters set as $e=2$, $k=2$, $v=2$, and as mentioned
above, $g=0$. The initial conditions are $q_{1}(0)=1,$ $q_{2}(0)=1,$
$q_{4}(0)=2,$ $p_{1}(0)=0,$ $p_{2}(0)=-1,$ $p_{3}(0)=0$ and $p_{4}(0)=0$.

The results are presented in a sequence of phase diagrams. In Figure 1, we display
$p_{4}$ versus $q_{4}$. We remind that we could have used any pair of variables
to perturb the initial fixed point and then present the outcome in these graphs, but our
choice was guided by the fact that both variables $q_{3}\left(  \phi\right)  $\ and
\ $q_{4}\left(  M\right)  $\ \ are present in the Higgs-type potential of our
model. After this, in the Figure 2, we show a graph of SALI as a function of
$q_{3}$.
\begin{center}
  \vspace{0.3cm}
{\par\centering
\resizebox*{0.90\textwidth}{!}{\rotatebox{0}{\includegraphics[
natheight=6.531100in,
natwidth=5.614400in,
height=4.5645in,
width=3.9271in
]{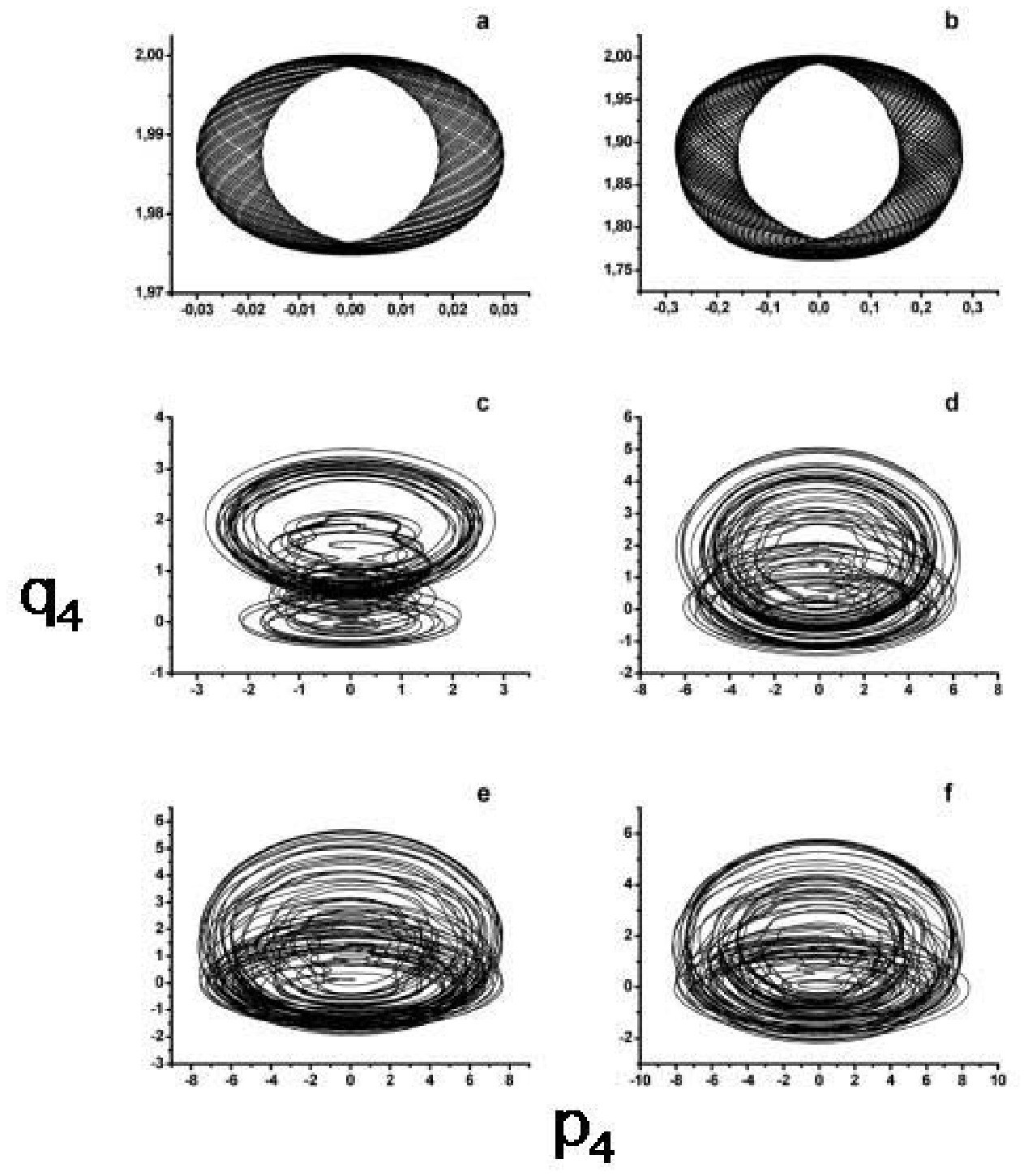}}}
\par}
{\par\centering  a)$q_{3}(0)=0.1$, b)$q_{3}(0)=0.3$, c) $q_{3}(0)=0.7$, d)$q_{3}(0)=1.5$, e) $q_{3}(0)=1.75$ and f) $q_{3}(0)=2.0$.\par}
{\par\centering Figure 1\par}
\vspace{1cm}

\end{center}

\bigskip

\begin{center}

  \vspace{0.3cm}
{\par\centering
\resizebox*{1.20\textwidth}{!}{\rotatebox{0}{\includegraphics[
natheight=13.280900in,
natwidth=17.187300in,
height=4.5619in,
width=5.8946in
]{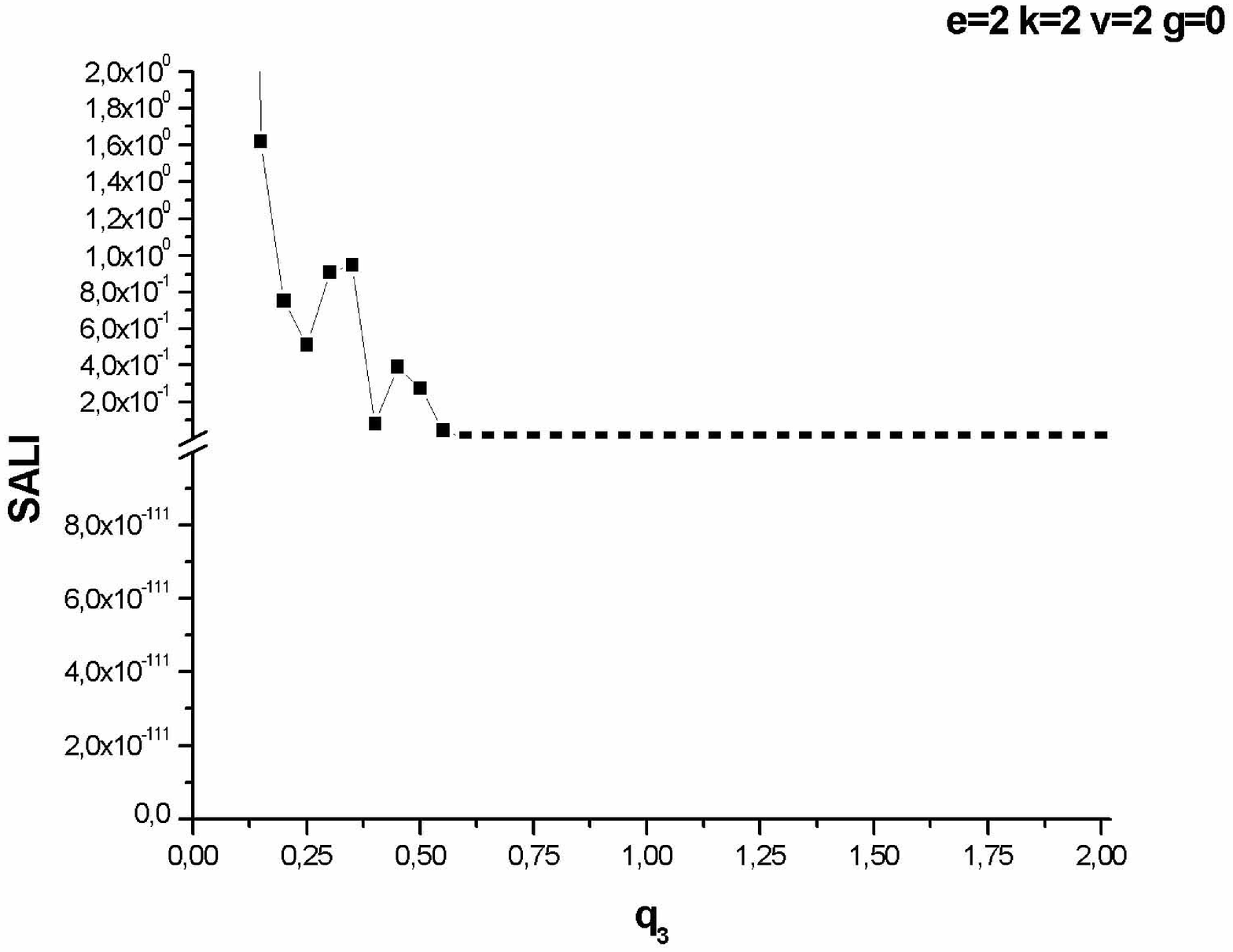}}}
\par}

{\par\centering Figure 2\par}
\vspace{1cm}

\end{center}

\bigskip

We can see from Figure 2, with the help of the break in the vertical axis, that for
$q_{3}\gtrsim0.55$ the behavior of the system becomes chaotic with SALI assuming
values between $10^{-89}$ and $10^{-33}$.
\bigskip

\subsection{\bigskip Case with g $\neq0$ but outside critical coupling regime}

\bigskip

Now, we use $g\neq0$, but outside critical coupling regime, to check whether the
inclusion of this kind of coupling may turn some configuration dynamics into
chaotic, with initial conditions that, in the case $g=0$,lead to regularity. To do that,
we fix the parameters and the initial conditions as $e=2$, $k=2$, $v=2$ and
$q_{1}(0)=1,$ $q_{2}(0)=1,$ $q_{3}(0)=0.25,$ $q_{4}(0)=2,$ $p_{1}(0)=0,$
$p_{2}(0)=-1,$ $p_{3}(0)=0$ and $p_{4}(0)=0$ and we analyze the model with
the following values of $g$: $0.1$, $0.7$, $1.5$ and $2.5$.

For all these cases, the behavior remained the same, indicating that the
variation of $g$ does not change the behavior of the system from regular to chaotic, as we can see in Figure 3.

\bigskip
\bigskip
\bigskip
\bigskip
\bigskip
\bigskip
\begin{center}

  \vspace{0.3cm}
{\par\centering
\resizebox*{0.70\textwidth}{!}{\rotatebox{0}{\includegraphics[
natheight=5.624700in,
natwidth=7.499600in,
height=5.6835in,
width=7.5688in
]{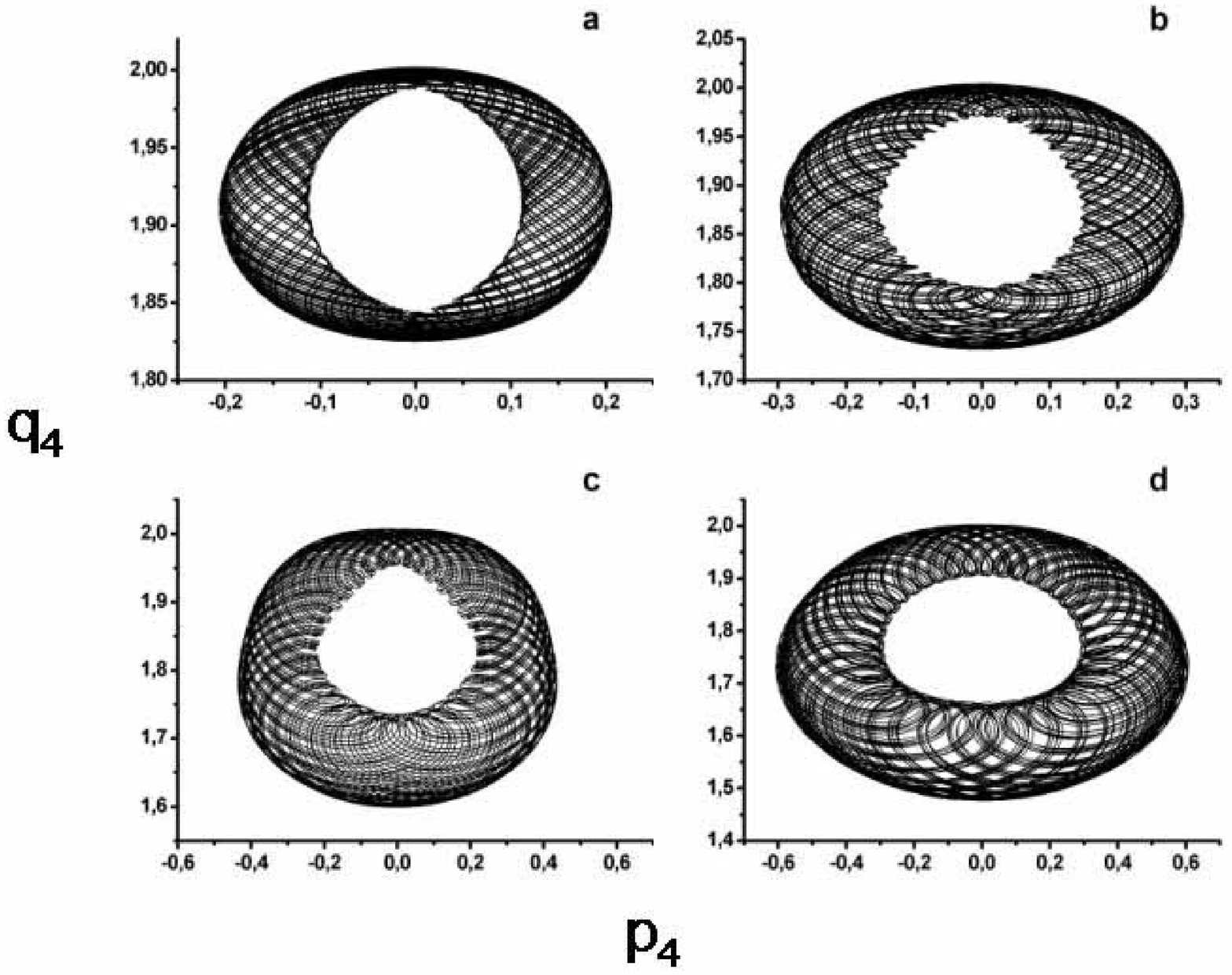}}}
\par}
{\par\centering  a)$g$=$0.1$,b) $g$=$0.7$,c) $g$=$1.5$ and d) $g$=$2.5$.\par}
{\par\centering Figure 3\par}
\vspace{1cm}

\end{center}

\bigskip

\subsection{Case with g in the critical coupling regime}

\bigskip

Now, we explore the critical coupling regime where $g=-\dfrac{e}{k}$. We fix
the parameters $k=2$, $v=2$ \ and vary $e$\ , keeping the initial
conditions as a perturbation case of the fixed point; but, in this case, we shall have
a different set of initial conditions for each $e$, since the general
expression for fixed point element p4(0) depends on $e$. With this in mind, we
keep the same values for $q_{1}(0)=1,$ $q_{2}(0)=1,$
$p_{1}(0)=0,$ $p_{2}(0)=-1,$ $p_{3}(0)=0,$ $p_{4}(0)=0$ and set $q_{3}(0)=0.7$,
a value that makes the system chaotic in the case $g=0$ and $q_{4}(0)=e$. In this case, $q_{3}(0)=0.7$ is the perturbation, since in the fixed point $q_{3}(0)=0.0$.
\bigskip As in the case g=0, we plotted a set of phase diagrams and a graph of $e$
versus SALI. In the SALI graphs, we break the vertical axis to show that the minimal SALI values are above the threshold of chaos, according to the expectations of the SALI method, that is, SALI $< 10^{-5}$.

\bigskip
\bigskip
\bigskip
\bigskip
\bigskip
\bigskip
\bigskip
\bigskip
\bigskip
\bigskip

\begin{center}

  \vspace{0.3cm}
{\par\centering
\resizebox*{0.60\textwidth}{!}{\rotatebox{0}{\includegraphics[
natheight=7.499600in,
natwidth=5.624700in,
height=4.5636in,
width=3.429in
]{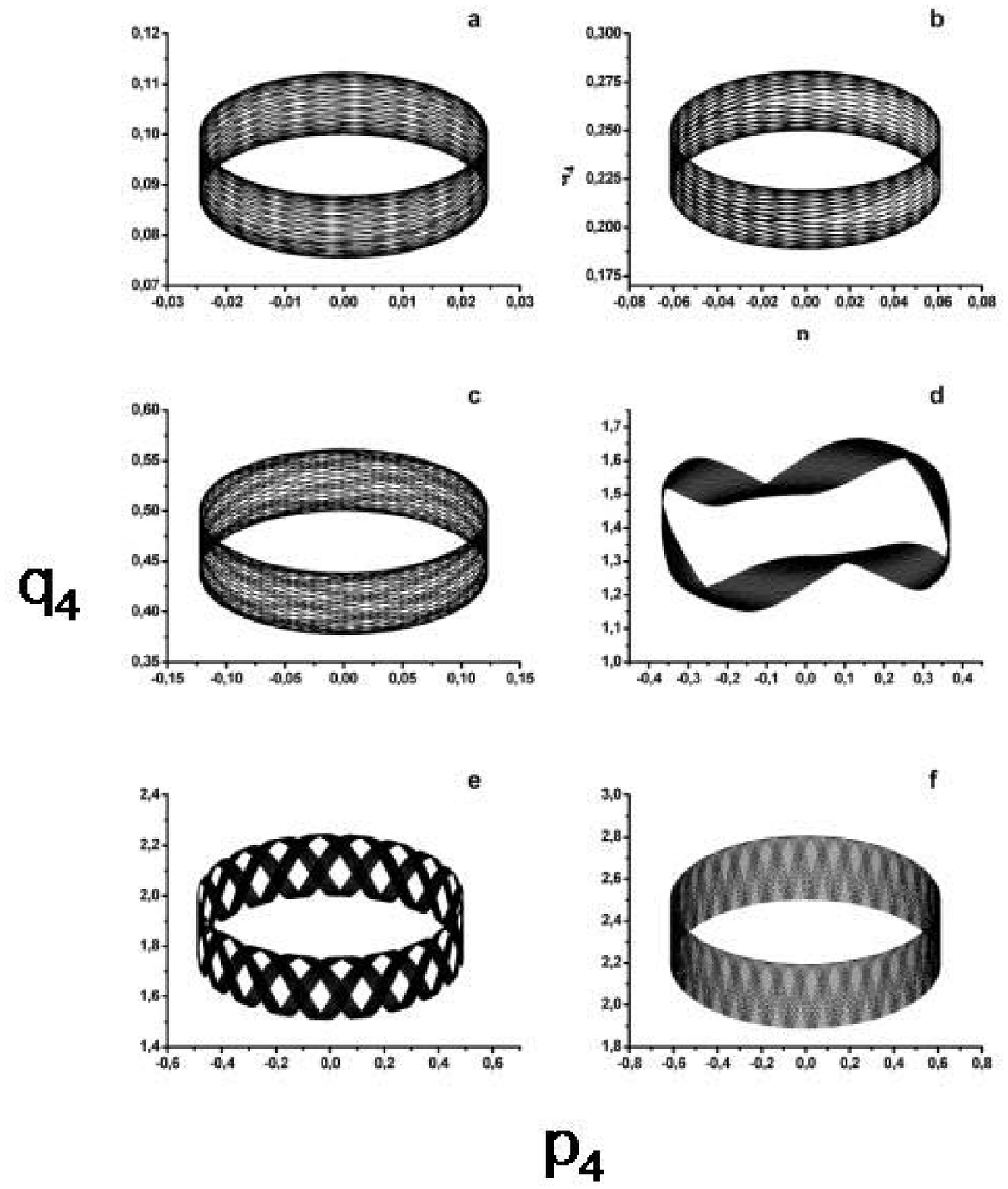}}}
\par}
{\par\centering $q_{3}(0)=0.7$ and $e$ varying as a)$e=0.10$, b)$e=0.25$, c) $e=0.50$, d)$e=1.5$, e) $e=2.0$ and f) $e=2.5$.\par}
{\par\centering Figure 4\par}
\vspace{1cm}

\end{center}
\bigskip

\begin{center}

  \vspace{0.3cm}
{\par\centering
\resizebox*{1.40\textwidth}{!}{\rotatebox{0}{\includegraphics[
natheight=13.280900in,
natwidth=17.187300in,
height=4.5619in,
width=5.8946in
]{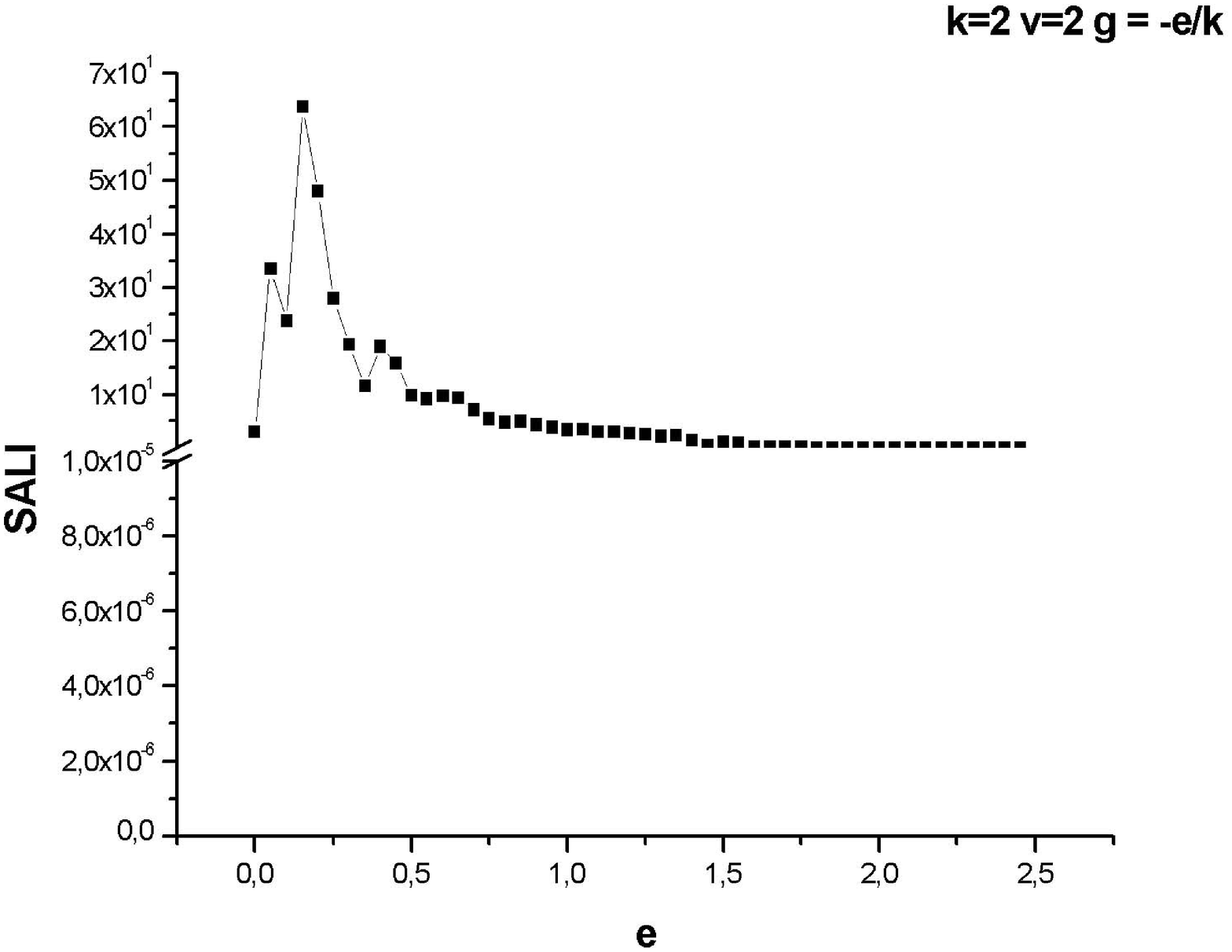}}}
\par}

{\par\centering Figure 5\par}
\vspace{1cm}

\end{center}

\section{Concluding comments}

The comparison encompassing the integrable critical coupling ``pure Chern-Simons''
system, presented in the work of Ref.\cite{mexico}, and our extended
N=2-\textit{susy} descent model, the latter deserving an integrability
analysis procedure both for the general and critical coupling regimes, as
reported here, suggests that the extra \textit{susy} may be responsible
\cite{intsusy} for the global non-integrability (in the strong Painlev\'{e} sense)
situation found even in the C.S.-like regime.

In studying chaos, two main possibilities have been checked. First, we have
verified if non-vanishing values of $g$ were able to drive the system (previously with
initial conditions and parameters such that regularity was achieved for $g=0$) into a chaotic regime.
The second point we have tackled concerns the opposite effect, namely, if a chaotic configuration for
$g=0$ may become regular whenever $g$ becomes non-trivial.

In the case of a non-critical coupling, as $g$ increases from zero, a stable configuration for $g=0$ keeps its stability as $g$ varies. In those cases, the difference is that, for larger values of $g$, the system becomes slightly more unstable, but its dynamics is still regular. Something similar happens for configurations that exhibit chaos for $g=0$. In such cases, the system remains chaotic, but a little more unstable.
For critical coupling, orbits that were chaotic for $g=0$ become now regular; this may indicate that the critical coupling plays the r\^ole of a stabiliser of our model.
These results may be interpreted on the basis of two points:

\begin{enumerate}
	\item For the critical coupling regime, there occurs a partial decoupling between the variables $\phi$ and $M$, and this reduces the non-linearity of the system.
	\item The quantity $G(G=1-g^{2}\phi^{2})$ must be positive, with $0<G\leq 1$. This must be so in order to ensure positivity of the energy, and the existence of a stable ground state.
	In Eqs.~(\ref{painleve1}),(\ref{painleve2}) and (\ref{painleve3}), $G$ accompanies all terms with time derivatives, and for large enough $g$ or $\phi$, $G$ becomes small, rendering the algebraic sectors of these equations dominant. This fact may have a stabilising consequence, implying that, in the case of a non-critical coupling, the dynamics for $g\neq0$ is not that different from the case with $g=0$. In the critical coupling regime, the stabilising effect could be a combination of the two reasonings just presented.  
\end{enumerate}
  
It is also noteworthy to point out that, for negative values of $G$, the phase space volume is no longer conserved, and, as a consequence, we do not have any longer a Hamiltonian system. For this reason, the results reported above make sense only if the initial conditions and the parameters ensure the positivity of $G$. Finally, as far as configurations with more regular behavior than the ones found in Ref.\cite{mexico}, as $g$ increases, show up in this model, we wonder whether special physical conditions - extended supersymmetry in the original system - might be responsible for this stabilising effect.

\bigskip \noindent{\bf Acknowledgments}\\

L.P.G.A., F.Haas and J.A. H.-N. express their gratitude to CNPq-Brazil for the invaluable financial support.
A.L.M.A. Nogueira thanks for FAT-UERJ, where the last part of this work has been performed.

%
%
%
%
%

%

\begin{thebibliography}{99}

\bibitem{chaology} B.A. Bambah, C. Mukku, M.S. Sriram, S. Lakshmibala,
hep-th/0203177 and references therein.

\bibitem{Chaos-gauge}T.S. Biro,S.G. Matinyan e B.M\"{u}ller, Chaos and Gauge
Field Theory (World Scientific Publishing Co Pte ltd, New Jersey, 1994).

\bibitem{savvidy} S.G. Matinyan, G.K. Savvidy, N.G.T. Savvidy,
Sov. Phys. JETP {\bf 53}, 421 (1981).

\bibitem{shur} E.S. Nikolaevskii, L.S. Shur, JETP Lett. {\bf 36}, 218
(1982).

\bibitem{bamba} B.A. Bambah, S. Lakshmibala, C. Mukku, M.S.
Sriram, Phys. Rev. {\bf D47}, 4677 (1993).

\bibitem{mexico} J. Escalona, A. Antill\'{o}n, M. Torres, Y.
Jiang, P. Parmananda, Chaos {\bf 10}, 337 (2000).

\bibitem{stern} J. Stern, Phys. Lett. {\bf B265 }, 119 (1991).

\bibitem{kogan} I.I. Kogan, Phys. Lett. {\bf B262}, 83 (1991).

\bibitem{navra} P. Navr\'{a}til, Phys. Lett. {\bf B365}, 119 (1996).

\bibitem{wit} E. Witten, D.I. Olive, Phys. Lett {\bf B78}, 97 (1978).

\bibitem{bogo} E.B. Bogomol'nyi, Sov. J. Nucl. Phys. {\bf 24}, 449 (1976).

\bibitem{antil} A. Antill\'{o}n, J. Escalona, M. Torres, Phys. Rev.
{\bf D55} 6327 (1997).

\bibitem{numero1} H.R. Christiansen, M.S.Cunha,
J.A.Helay\"{e}l-Neto, L.R.U. Manssur and A.L.M.A. Nogueira, Int.
J. Mod. Phys. {\bf A14} 147 (1999).

\bibitem{selfdual} H.R. Christiansen, M.S.Cunha,
J.A.Helay\"{e}l-Neto, L.R.U. Manssur and A.L.M.A. Nogueira, Int.
J. Mod. Phys. {\bf A14} 1721 (1999).

\bibitem{sarlet} Sarlet, W. \& Cantrijn, F. \, {\it Generalizations of Noether's Theorem in Classical Mechanics}.
\, SIAM Rev. {\bf 23} 467-494 (1981).

\bibitem{olver} Olver, P. J. \, {\it Applications of Lie Groups to Differential Equations}.
Graduate Texts in Mathematics No. 107. \, Springer-Verlag: New York, 1986.


\bibitem{pain1}M.Tabor, Chaos and Integrability in Non-Linear Dynamics : An
Introduction (John Wiley \& Sons, Inc., New York, 1989).

\bibitem{pain2}M.J. Ablowitz, A. Ramani, H. Segur, Lett. Nuovo Cim.
\textbf{23} (9),(1978) 333.


\bibitem{lya1}G.Benettin, L. Galgani, A. Giorgilli, J.M. Strelcyn, Meccanica \textbf{15}, 21 (1980).

\bibitem{lya2}A. Wolf,J.B. Swift, H.L. Swinney, J.A. Vastano, Physica D \textbf{16}, 285 (1985).


\bibitem{SABV1} Skokos, C; Antonopoulos, C; Bountis, TC; Vrahatis, MN., Detecting order and chaos in Hamiltonian systems by the SALI method, J. Phys. A-Math. Gen. {\bf37}, 6269 (2004) .

\bibitem{SABV2} Skokos, C; Antonopoulos, C; Bountis, TC; Vrahatis, MN,  How does the Smaller Alignment Index (SALI) distinguish order from chaos?, Prog. Theor. Phys. Suppl. (150), 439-443 2003.

\bibitem{intsusy} As an example of ``susy-spoiling"  of integrability
we suggest J.M. Evans, J.O. Madsen, Phys. Lett. {\bf B389}, 665 (1996) .




\end{thebibliography}
\end{document}